\documentclass[12pt]{iopart}   
\usepackage{iopams}
\usepackage{epsf}
\usepackage{euscript}
\usepackage{graphicx}

\renewcommand{\footnoterule}

\begin{document}
\title {Resolving isospectral ``drums" by counting nodal domains}
\author{Sven Gnutzmann$^2$ \footnote{Presently on leave at the
Department of Physics of Complex Systems,\\ The Weizmann Institute
of Science, Rehovot 76100, Israel}, Uzy Smilansky$^1$\footnote
{Presently on sabbatical leave at the School of Mathematics,
Bristol University, Bristol BS81TW, UK} and Niels
Sondergaard$^1$\footnote{ Present address: Division of
Mathematical Physics, Lund Institute of Technology Lund University
Box 118 SE - 221 00 Lund SWEDEN}}
\address {$^1$ Department of Physics of Complex Systems,\\ The Weizmann
Institute of Science, Rehovot 76100, Israel}
\address{$^2$ Institute for Theoretical Physics,\\
Freie Universit\"{a}t Berlin, 14195 Berlin, Germany.}
\date{\today}

\begin{abstract}

Several types of systems were put forward during the past decades
to show that there exist {\it isospectral} systems which are {\it
metrically} different. One important class consists of Laplace
Beltrami operators for pairs of flat tori in $\mathbb{R}^n$ with
$n\geq 4$. We propose that the spectral ambiguity can be resolved
by comparing the nodal sequences (the numbers of nodal domains of
eigenfunctions, arranged by increasing eigenvalues). In the case
of isospectral flat tori in four dimensions - where a 4-parameters
family of isospectral pairs is known-  we provide heuristic
arguments supported by numerical simulations to support the
conjecture that the isospectrality is resolved by the nodal count.
Thus - one can {\it count} the shape of a drum (if it is designed
as a flat torus in four dimensions\dots).
\end{abstract}

\section{Introduction}
\label{sec:introduction} Since M. Kac posed his famous question:
``can one hear the shape of a drum" \cite{kac}, the subject of
isospectrality appears in many contexts in the physical and the
mathematical literature. This question can be cast in a more
general way by considering a Riemannian manifold (with or without
boundaries) and the corresponding Laplace-Beltrami operator.
(Boundary conditions which maintain the self adjoint nature of the
operator are assumed when necessary.) Kac's question is
paraphrased to ask ``can one deduce the metric of the surface (or
the geometry of the boundary) from the spectrum?". Till today, the
answer to this question is not known in sufficient detail. An
affirmative answer is known to hold for several classes of
surfaces and domains \cite{Bruening84,Zelditch98,gutkinus} (see
also a recent review by S. Zelditch \cite {Zelditch04}). However,
this is not always true. One of the first examples to the negative
is due to J. Milnor who proposed in 1964 two flat tori in
$\mathbb{R}^{16}$, which he proved to be isospectral but not
isometric \cite{Milnor}. Since then, many other pairs of
isospectral yet not isometric systems were found. M. E. Fisher
considered a discrete version of the Laplacian, and gave a few
examples of distinct graphs which share the same spectrum
\cite{Fisher}. A general method for constructing isospectral,
non-isometric manifolds has been designed by T. Sunada
\cite{Sunada}. Sunada's technique applies also to discrete graphs
\cite{Brooks}, and the late Robert Brooks \cite{Brooks}, gave a
few examples of families of non-Sunada discrete graphs. Sunada's
method was also used by Gordon {\it et. al.} \cite{Gordon} and
Chapman \cite {chapman} to construct isospectral domains in $
\mathbb{R} ^2$. Other pairs of isospectral domains in $ \mathbb{R}
^2$ were proposed in \cite {Buser} and discussed further in
\cite{Okada}. Sunada-like quantum graphs were presented in
\cite{gutkinus}.

Milnor's original work on isospectral flat tori in
$\mathbb{R}^{16}$ induced several investigators to find other
examples in spaces of lower dimensions. Kneser \cite{Kneser}
constructed an  example in dimension $n=12$, and proved that there
exist no such pair in two dimensions. Wolpert \cite{Wolpert}
showed that all sets of mutually isospectral but non-isometric
flat tori are finite at any dimension.  The first examples in
dimension $n=4$ were found by Schiemann \cite {Schiemann}, and
later by Earnest and Nipp \cite{Earnest}. These results were
generalized by Conway and Sloane \cite{Conway}, who constructed
sets of isospectral pairs of flat tori in $n=4,5,6$, and these
sets depend continuously on several parameters.

The existence of such a large variety of isospectral pairs,
suggests naturally the question - what is the additional
information necessary to resolve the isospectrality. We would like
to propose that this information is stored in the sequences of
nodal counts, defined as follows. Consider only real
eigenfunctions of the Laplace-Beltrami operator and assign to each
eigenfunction the number of its connected domains where the
eigenfunction does not change its sign (such a domain is a nodal
domain). The nodal sequence is obtained by arranging the number of
nodal domains by the order of increasing eigenvalues. Sturm's
oscillation theorem in one dimension, and Courant's generalization
to higher dimensions express the intimate relation between the
nodal sequence and the corresponding spectrum. However, the
information stored in the nodal sequence and in the spectrum is
not the same, and here we would like to propose that the
additional knowledge obtained from the nodal sequence can resolve
isospectrality. We address in particular isospectral flat tori in
four dimensions of the type mentioned above. The simplicity of the
geometry, together with the rich variety of pairs, make this class
of systems very convenient, especially here, when the approach is
explored for the first time.

The paper is organized as follows. In  section (\ref {sec:flat
tori}) we shall summarize some of the properties of flat tori,
their spectra and eigenfunctions. The fact that the spectra are
highly degenerate requires a special choice of the basis set of
wave functions for which the nodal domains are to be counted. The
quantity which signals the difference between the isospectral tori
is defined in subsection (\ref {subsec:count}). The arguments
which lead us to suggest that this quantity resolves
isospectrality are explained for the families of flat tori in four
dimensions \cite{Conway}, and it is presented in subsection (\ref
{subsec:dist}). A summary and some concluding remarks will be
given in the last section.

\section{Flat Tori}
\label{sec:flat tori}

A flat torus is a Riemannian manifold which
is a quotient of $\mathbb{R}^n$ by a lattice of maximum rank:
$T=\mathbb{R}^n/A\mathbb{Z}^n$, where $A=(\bf{g}^{(1)},\cdots \bf
{g}^{(n)})$. Thus, the lattice $A\mathbb{Z}^n$ is spanned by the
$\bf{g}^{(r)}$. The reciprocal lattice will be denoted by $\hat
 {\bf g}^{(r)}$, and $ (\hat{\bf g}^{(s)}\cdot{\bf
g}^{(r)}) = \delta_{r,s}$. The Gram matrices for the lattice will
be denoted by $G=A^\top A$, and its reciprocal will be denoted for
brevity by $Q=G^{-1}=(A^{-1})(A^{-1})^\top $. In the present work
we shall deal with dimensions $n\ \ge\ 4$. We shall assume
throughout that the lattice vectors $A$ cannot be partitioned to
mutually orthogonal subsets.

\subsection{Spectra}
\label{subsec:spect} We are interested in the spectrum of the
Laplace Beltrami operator $\Delta =
-\sum_{i=1}^n\frac{\partial^2}{\partial x_i^2}$ with
eigenfunctions which are uniquely defined on $T$. They can be
explicitly written down as:
\begin{equation}
\Psi_{\bf q}({\bf x})= \exp \left (2\pi{\rm i}\sum_{r=1}^n q_r
(\hat{\bf g}^{(r)}\cdot{\bf x})\right ) \label{eq:cwfunct}
\end{equation}
where ${\bf q}=(q_1,\cdots,q_n) \in \mathbb{Z}^n , {\bf x}\in T$.
The corresponding eigenvalues are
\begin{equation}
E_{\bf q}=(2\pi)^2({\bf q}\cdot Q {\bf q}) \label{eq:spect}.
\end{equation}
The spectrum of a flat torus may be degenerate, and we denote the
degeneracy by
\begin{equation}
  g_Q(E)= \sharp \left \{{\bf q}\in \mathbb{Z}^n : E= E_{\bf q}\right \}\
  .
  \label{eq:degenracy}
\end{equation}
If the matrix elements of $Q$ are {\it rational}, it is convenient
to express the energy in units of $(2\pi)^2/l(Q)$ where $l(Q)$ is
the least common denominator of the elements of $Q$. In these
units the energy values are integers, and $g_Q(E)$ equals the
number of times that $E$ can be represented as an integer
quadratic form. The integer vectors which satisfy (\ref
{eq:spect}) will be called {\it representing vectors} in the
sequel. Their tips are points on the $n$ dimensional ellipsoid
(\ref {eq:spect}), and their distribution on the ellipsoid will be
discussed in the sequel. The study of the spectrum
(i.e. those integers that can be represented by a given
integer quadratic form)
and the degeneracies
(the number of representations) is a subject which was studied at
length in number theory. Here, we shall give a brief summary of
the results which are essential for the present work. The
interested reader is referred to \cite{Cassels,Iwaniec} for
further references and details.

\noindent {\it (a)}
For integer $Q$, the eigenvalues are integers
$E$ which must satisfy the congruence
\begin{equation}
  E = m (\ {\rm mod}\  c(Q)\ ) \ \ ; \ \ m \in \mathbb{N}
  \label{eq:congruence}
\end{equation}
and $c(Q)$ is an integer which depends on $Q$. This result implies
that the spectrum is {\it periodic}.

\noindent {\it (b)}
The degeneracy $g_Q(E)$ for {\it integer} $Q$
increases as $E^{\frac{n}{2} -1}$.
This estimate can be derived by
a simple heuristic argument. The number of integer grid points in
a shell of width $\delta E$ is proportional to $E^{\frac{n-1}{2}}
\delta E/  E^{\frac{1}{2}} $. If $Q$ is integer, $E$ must be an
integer, and hence, the number of values it can take in the
interval of interest is $ \delta E$. Thus,
\begin{equation}
  g_{rat}(E) \sim \frac{\sharp\ {\rm grid\ points}}{\sharp\
    {\rm possible \ values}} \propto E^{\frac{n}{2}-1}\ .
  \label{eq:degest1}
\end{equation}
From {\it (a)} and {\it (b)} above, it follows that not all
integers appear in the spectrum, and the gaps are determined by
the gaps which appear in the first $c(Q)$ eigenvalues in the
spectrum. The distribution of the gap sizes may be very complex,
and depend delicately on $Q$. Still, one can define the mean gap
size, and this value applies for the entire spectrum.

\noindent {\it (c)}
For {\it irrational} matrices of the form $Q=Q_0+\alpha Q_1$ where
$Q_0,Q_1$ are both integer, and $\alpha$ is irrational, the mean
degeneracy increases more slowly with $E$, namely,
$g(E) \propto E^{\frac{n}{2}-2}$. Here, the degeneracy
class consists of the grid points which satisfy both $E_0  = {\bf
q}.Q_0 {\bf q}$ and $E_1 ={\bf q}.Q_1 {\bf q}$ with $ E=E_0  +
\alpha E_1  $, and both $E_0$ and $E_1$ integers. Their number is
proportional to the volume of the intersection of the
corresponding ellipsoid shells
$ E^{\frac{n-2}{2}}  \delta E_0
 \delta E_1/ (E_0E_1)^{\frac{1}{2}}$.
The number of
spectral values is now the number of integer points in the square
of size $ \delta E_0 \delta E_1 $. Thus, in the irrational case,
\begin{equation}
  g_{irrat}(E) \propto E^{\frac{n}{2}-2}\ .
  \label{eq:girrat}
\end{equation}

Figure 1. shows the dependence of the mean value of $g(E)$,
averaged over the eigenvalues $E_i$ in an interval of width
$\Delta E$ centered at $E$,
 \begin{equation}
\langle g(E) \rangle\ =\ \frac{1}{M} \sum_{i: |E_i-E|<\Delta E/2}
g(E_i) \ . \label{eq:spect-av}
\end{equation}
$M$ stands for the number of eigenvalues in the interval. The
linear dependence of $\langle g(E) \rangle$ at sufficiently large
$E$ for rational $Q$ and its independence of energy in the
irrational case are clearly illustrated.

\begin{figure}
  \begin{center}
    \includegraphics[width=0.8\linewidth]{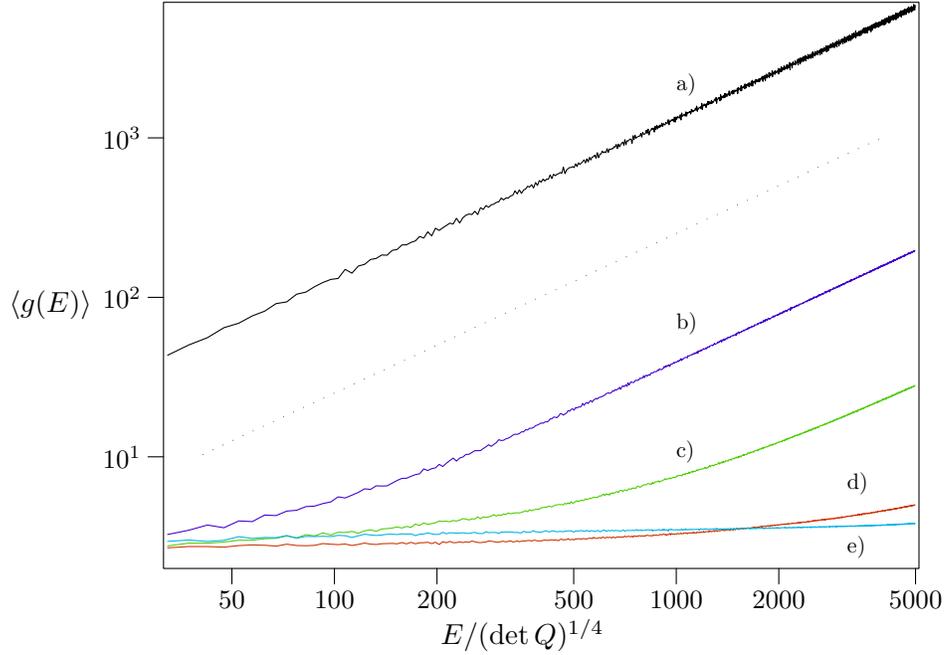}
    \caption{
      Double logarithmic plot of the
      spectral averaged degeneracy (\ref{eq:spect-av}).
       The five data sets correspond to:
      a) $\beta=2$, b) $\beta=3/2$, c) $\beta=5/3$, d) $\beta=8/5$
      and e) $\beta=(\sqrt{5}+1)/2$. a)-d) correspond to integer tori
      and e) to an irrational. The parameter $\beta$ is explained
      in the text at the end of section \ref{subsec:def}.
      The dotted line is a linear function of $E$ included
      for comparison.
    }
    \label{figure1}
  \end{center}
\end{figure}

\noindent {\it (d)} For integer $Q$, $n\ge 4$ and in the limit of
large $E$, the representing integer vectors of $E$ are uniformly
distributed on the ellipsoid: Given a well behaved function
$f({\bf x})$ on the unit ellipsoid $\mathcal{E}\ =\left \{ {\bf x}
\in \mathbb{R}^n : {\bf x}\cdot Q{\bf x}=1\right \}$,
\begin{equation}
 \lim_{E\rightarrow\infty}\ \frac{1}{g(E)} \ \sum_{{\bf q}\cdot Q {\bf q}=E} f(\frac {{\bf
q}}{\sqrt{E} }) - \int_{\mathcal{E}}f({\bf x}){\rm d}{\bf x} = 0.
 \label {eq:ergdc}
\end{equation}

\subsection {Isospectral flat tori in four dimensions}
\label{subsec:def} To end this section, we shall describe the
4-parameter family of non isometric yet isospectral flat tori
which is the system we shall consider in most of the examples to
be discussed in the sequel. This family which was discovered by J.
H. Conway and N. J. A. Sloane \cite {Conway} includes, as a
particular case, the previously known example of Schiemann \cite
{Schiemann}. Their spectra are given explicitly by the pairs of
positive definite matrices, which depend on the 4 positive
parameters $a,b,c,d$:
\vspace{0.5cm}
\vbox{\small
\begin {eqnarray} \label{eq:defineQ}
\fl  Q^{+} =
  \frac{1}{12}\left (
    \begin{array}{cccc}
      9a + b + c + d & 3a - 3b -  c + d &3a +  b - 3c - d &3a -  b +  c - 3d \\
      3a -3b - c + d &  a + 9b +  c + d & a - 3b + 3c - d & a + 3b -  c - 3d \\
      3a + b -3c - d &  a - 3b + 3c - d & a +  b + 9c + d & a -  b - 3c + 3d \\
      3a - b + c -3d &  a + 3b -  c -3d & a -  b - 3c +3d & a +  b +  c + 9d
    \end{array}
  \right ) \mbox{\normalsize \, and} \\ \nonumber \\ \nonumber
\fl  Q^{-} =
  \frac{1}{12}\left (
    \begin{array}{cccc}
      9a + b + c + d & -3a + 3b - c + d &-3a + b + 3c - d &-3a - b + c +3d \\
      -3a +3b - c + d & a + 9b + c + d & a + 3b - 3c - d & a - 3b - c + 3d \\
      a + b + 3c - d &a + 3b - 3c - d & a + b + 9c + d & a - b + 3c - 3d  \\
      -3a - b + c +3d & a - 3b - c + 3d & a - b + 3c - 3d & a + b + c +9d
    \end{array}
  \right )  \, . 
\end{eqnarray} }
Several properties of these matrices can be derived by straightforward computations:

\noindent {\it i.} The spectra of  $Q^{+}$ and $Q^{-}$ are
identical, and consist of the values $a,b,c,d$. The unitary
matrices which bring $Q^{\pm}$ to diagonal form are {\it
  independent} of the parameters. Explicitly

\begin{eqnarray}  \label{eq:tpm}
D&\equiv &{\rm diag} \{a,b,c,d \}= T^{\pm}\ Q^{\pm}\ (T^{\pm})^{\top}  \\ \nonumber  
\\   \small
T^{\pm}&=&
  \frac{1}{  \sqrt{12}}\left[ \left (
      \begin{array}{rrrr}
        0 & 1 &  1 & 1  \\
        -1 &0 & -1 & 1  \\
        1 & -1 & 0 & 1  \\
        1 & 1 &  -1 &0
      \end{array}
    \right ) \pm  3 \left(
      \begin{array}{rrrr}
        1 & 0 &  0 & 0  \\
        0 & 1 & 0 & 0  \\
        0 & 0 & -1 &0  \\
        0 &0 &  0 &-1
      \end{array}
    \right ) \right] \, .  \nonumber \normalsize
\end{eqnarray}

\noindent {\it ii.} $Q^{+}$ and $Q^{-}$  commute only when at
least three of the parameters $a,b,c,d$ are equal.  From now on
when we shall refer to the four parameters family of isospectral
tori, we shall exclude this set which represents ellipsoids with
cylindrical or spherical symmetry.

\noindent {\it iii.} The unitary matrix $U$ which transform
$Q^{+}$ to $Q^{-}$,  $Q^{-}=U^{\top} Q^{+}U$ is {\it independent}
of the parameters $a,b,c,d$,
\small
\begin {eqnarray}
  \label{eq:utu}
  U =(T^{+})^\top\ T^{-} =
  \frac{1}{2}\left (
    \begin{array}{rrrr}
      -1 & 1 &  1 & 1  \\
      -1 &-1 & -1 & 1  \\
      -1 & 1 & -1 &-1  \\
      -1 &-1 &  1 &-1
    \end{array}
  \right ) \ \ \ \ ; \ \ \ \det U =1\ .
\end{eqnarray}
\normalsize

Consider an integer vector ${\bf q}$ with ${\bf q}\cdot Q^{-}{\bf
  q}=E$. $U{\bf q}$ is on the ellipsoid generated by $Q^{+}$,  but
it is \emph{integer} only if $\sum_{i=1}^4 |q_i|$ is even. In this
case, also $\sum_{i=1}^4|\left(U{\bf q}\right)_i|$ is even. Hence,
integer vectors with \emph{even sums} map to each other under $U$.
The integer vectors with \emph{odd sums} do not have this
property.

Conway and Sloane's paper offers another family of isospectral
tori in four dimensions. This is a two parameter family which we
shall not discuss in detail, although the numerical results
obtained for this case support the conclusions derived from the
study of the four parameters family.

To end this section we would like to describe the numerical
simulations which accompany the subsequent discussions.
We have calculated the first 120 million eigenvalues for
five pairs of isospectral tori (together with the corresponding
nodal sequences defined in the next section). Four of the
pairs have been chosen rational, one irrational.
 The four parameters that define a
pair (see eq. (\ref{eq:defineQ})) are all of the form
$(a,b,c,d)=(\alpha, \alpha/\beta^2, \alpha/\beta^4,
\alpha/\beta^6)$. Here $\alpha$ just rescales the spectrum, for
the rational tori it is taken such that the tori are actually
integer. For the irrational pair it has been set to $\alpha=1$.
Note, that we present all results as a function of
$E/(\mathrm{det}\, Q)^{1/4}$ which is invariant under rescalings
of the spectrum. For the other parameter we have chosen the five
values $\beta_i=2,3/2,5/3,8/5,(\sqrt{5}+1)/2$. The latter value
defines the irrational pair ($\beta_5= (\sqrt{5}+1)/2$ is the
golden ratio), the other parameters are rational approximants to
the golden ratio along the Fibonacci sequence. Integer tori are
obtained by setting $\alpha_i=2^8,2^2\times 3^7,3\times
5^6,2^{20}$ (which results in
$\mathrm{det}\,Q(\alpha_i,\beta_i)=2^{20},3^{16}\times
2^{20},3^{16}\times5^{12},2^{44} \times 5^{12}$).

\section{Nodal domains and isospectrality}
\label{sec: nodal}
In this section we shall define the nodal
sequences of the flat tori under consideration, and will show how
they can resolve isospectrality.

To define the nodal domains, we consider the real counterparts of
(\ref {eq:cwfunct}),
\begin{equation}
\hspace {-1cm}
 \Psi^{(+)}_{\bf q}= \cos \left (2\pi\sum_{r=1}^n q_r (\hat{\bf g}^{(r)}\cdot{\bf x})\right ) \ \ ; \
\ \Psi^{(-)}_{\bf q}= \sin \left (2\pi\sum_{r=1}^n q_r (\hat{\bf
g}^{(r)}\cdot{\bf x})\right ) \label{eq:rwfunct}
\end{equation}
and to avoid double counting, we must exclude $-{\bf q}$ if ${\bf
  q}$ is included.

A convenient representation can be obtained by the transformation
\begin{equation}
  {\bf y} = G{\bf x} \ \  ;\ \ {\bf x}\in T \ \ ,\ \ {\bf y}\in
  T_y=\mathbb{R}^n/\mathbb{Z}^n
\end{equation}
so that
\begin{equation}
  \Delta_y  =  -\sum_{r,s=1}^n Q_{r,s}\frac{\partial^2}{\partial
    y_r\partial y_s}
\end{equation}
and
\begin{equation} \Psi^{(+)}_{\bf q}({\bf y}) =\cos\left(2\pi ({\bf
      q}\cdot {\bf y})\right) \ \ ; \ \ \Psi^{(-)}_{\bf q}({\bf y})
  =\sin\left(2\pi ({\bf q}\cdot {\bf y})\right) \ .
  \label{eq:realwf}
\end{equation}
From this representation it is evident that as long as the lattice
vectors cannot be reduced to mutually orthogonal subspaces, the
{\it real} wave functions cannot be expressed in a product form.
The number of nodal domains is defined by lifting the wave
functions to $\mathbb{R}^n$ and counting their nodal domains in
the unit cell. The nodal domains form parallel strips separated by
$n-1$ dimensional nodal hyper-planes. The  number of nodal domains
is
\begin{equation}
  \hat{\nu}[{\bf q}] = 2\sum_{r=1}^n |q_r|\ .
  \label{eq:nutorus}
\end{equation}
This result can be proved by induction, and we start by assuming
that all the $q_i$ are positive. (\ref {eq:nutorus}) is trivially
true for $n=1$, where the nodal manifold are points. In $2$-d, the
nodal manifold are lines which are perpendicular to the direction
$(q_1,q_2)$. There are $2q_1$ such lines which intersect the unit
interval on the $y_1$ axis at the points which are the nodal
manifolds of the $1$-d problem. There are additional $2q_2-1$
lines which intersect the interval $1<y_2<1$ on the line $y_1=1$.
In total there are $2(q_1+q_2)-1$ nodal lines in the unit interval
and therefore $2(q_1+q_2)$ nodal domains. The same argument can
now be repeated for any $n$. The case when some $q_i$ are negative
can be taken care of by a proper reflection. It is also clear that
the number of nodal domains of any linear combination of
$\Psi^{(+)}_{\bf q}({\bf y})$ and $\Psi^{(-)}_{\bf q}({\bf y})$
(\ref {eq:rwfunct}) depends on ${\bf q}$ only.

The strip structure discussed above is very different from the
morphology of nodal domains in \emph{separable} systems, which are
formed from intersections of locally perpendicular hyper-planes.

The special feature of the flat tori to be discussed here is that
their spectra are degenerate, and the degeneracies are maximal
when the tori are integer. Any linear combination of the
eigenfunctions in the degenerate space is itself an eigenfunction,
and its nodal structure may depend on the particular combination.
To be definite, we must chose a unique representation of the wave
functions, for which the nodal count can be defined in an
unambiguous way. For this purpose we require that the
eigenfunctions are presented in the form of a single term and the
nodal domains are parallel strips in the unit cell. This
definition singles out a well defined basis for each degeneracy
space, and the number of nodal domains is given by (\ref
{eq:nutorus}).

\subsection {Counting nodal domains}
 \label{subsec:count}
In a previous work \cite {BGS}, we have shown that counting nodal
domains can reveal important information concerning the classical
dynamics of the system under study. In particular, arranging the
spectrum by increasing values of the eigenvalues, we studied the
distribution of the normalized nodal counts $\xi_n=\frac
{\hat{\nu}_n}{n}$, where $\hat{\nu}_n$ is the number of nodal
domains of the $n$'th wave function. Courant's theorem guaranties
that $\xi_n \le 1$. The $\xi$ distributions have characteristic
and universal form if the problem is separable, which is distinct
from the distribution which prevails when the classical dynamics
is chaotic.

As long as the degeneracy in the spectrum is accidental, one can
ignore the ambiguity induced in the ordering of the spectrum by
the degeneracy. In the present case, the degeneracy increases as a
power of $E$, and therefore a different quantity, which is
independent of the ordering of the wave functions within a
degeneracy subspace, is called for.

A natural modification of the normalized nodal counts for a highly
degenerate spectrum is obtained by considering the quantity
\begin{equation} \hspace{-1cm}
\nu_{Q}(E) = \frac{1}{g_Q(E)}\sum_{{\bf q}\in \mathbb{Z}^n : E=
{\bf q}\cdot Q{\bf q}} \hat{\nu} [{\bf q}]\ \ =\ \
\frac{1}{g_Q(E)}\sum_{{\bf q}\in \mathbb{Z}^n : E= {\bf q}\cdot
Q{\bf q}}   [\ 2 {\small \sum_{i=1}^n} |q_i| \  ]\ .
 \label{eq:modnod}
 \end{equation}
We shall refer to $\nu_{Q}(E)$ as the {\it nodal count} associated
with the degenerate eigenvalue $E$. It coincides with the standard
definition for non degenerate cases. The nodal count is the tool
by which we propose to resolve isospectrality. It is defined as an
average over the representing vectors on the surface of the
ellipsoid $E= {\bf q}\cdot Q{\bf q}$. Because of (\ref
{eq:ergdc}), the mean value of  $\nu_{Q}(E)$ for large $E$ assumes
a well defined, and geometrically appealing value.

\subsection {Nodal counts and isospectrality}
\label{subsec:dist}

In this section we shall summarize the evidence we have in support
of the conjecture that the sequences of nodal counts of tori which
are isospectral but not isometric are different. Denoting by $Q^+$
and $Q^-$  the corresponding pair of Gramm matrices, and by $E$ a
spectral point, we study the difference
\begin{equation}
 \delta  \nu (E) \  =\  \nu_{Q^{+}}(E)- \nu_{Q^{-}}(E)\ .
 \label{eq:deltanu}
\end{equation}
Rather than examining (\ref {eq:deltanu}) for individual
eigenvalues, we shall consider its average over spectral
intervals, and the fluctuations about the mean.

Using the fact that in the limit of large $E$ the representation
vectors are distributed homogeneously over the ellipse  (\ref
{eq:ergdc}), we get,
\begin{eqnarray}
\left \langle \nu_Q(E)\right \rangle &\propto& E^{\frac{1}{2}}
\int_{\mathbb{R}^4} {\rm d}{\bf s}\ \delta(1-{\bf s}\cdot Q{\bf
s})  \hat{\nu}[{\bf s}]
\nonumber \\
 &=& E^{\frac{1}{2}} \int_{\mathbb{S}^3}
 \frac{{\rm d}\bomega \ \hat{\nu}[\bomega] }{{(\bomega \cdot Q \cdot
\bomega)}^{5/2}} \ ,
 \label {eq:nuav}
\end{eqnarray}
where the factor $E^{\frac{1}{2}}$ is due to the linear dependence
of $\hat \nu [{\bf q}]$ on ${\bf q}$.  We use this expression to
prove that $ \left \langle\delta \nu (E)\right \rangle =0 $. This
is trivially true for $a=b=c=d$, where the two ellipsoids
degenerate to a sphere.  In \ref {appendix1} we show this in
general, i.e. the smooth nodal counts agree. The numerical
simulations reproduce this behavior, and we use this fact as a
check of the numerical procedure.

The fluctuations in the difference of the nodal sequences are best
demonstrated by studying the variance
\begin{equation}
\hspace {-2.5cm}   \langle(\delta \nu(E))^2 \rangle  = \frac{1}{M}
\sum _{|E-E_m|<\Delta/2} \left (
\nu_{Q^+}(E_m)-\nu_{Q^-}(E_m)\right )^2 \ \ ;\ E_m {\rm \ in\ the\
spectrum.}
 \label{eq:variance}
\end{equation}
Here $\Delta \ll E$ and  $\ M=\sharp \{E_m:|E-E_m|<\Delta/2\} $.
The variance $\langle(\delta \nu(E))^2 \rangle$ was computed for
the examples of non isometric yet isospectral tori discussed in
conjunction with figure 1.  The results are shown in Figure 2. The
variance does not vanish, and its $E$ dependence clearly
distinguishes between the rational and irrational pairs. It shows
the variance for the isospectral tori which were introduced at the
end of section (\ref{subsec:def}). The two extreme values
($\beta=2$ and $\beta = ({\sqrt 5} +1)/2$) show the expected
dependence on $E$ over the entire range of $E$ values. The data
for the intermediate sets of parameters follow  the expected
asymptotic behavior  only after $E$ is large enough to
``distinguish" that the coefficients are rational. The same
behavior was also observed for the two parameter family of
isospectral tori in four dimensions mentioned in \cite{Conway}.

\begin{figure}
  \begin{center}
    \includegraphics[width=0.8\linewidth]{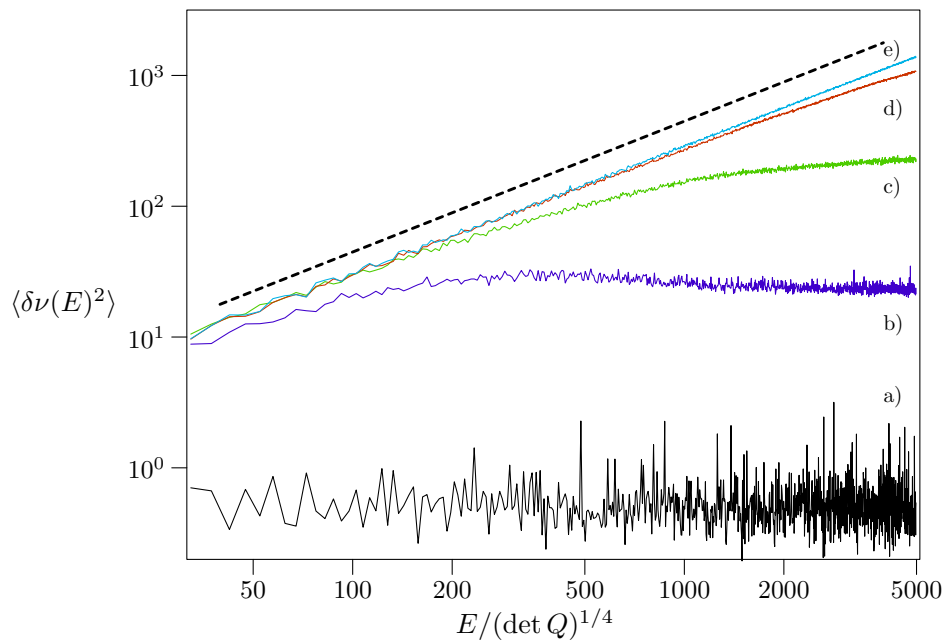}
    \caption{Double logarithmic plot of the variance
      $\langle \delta\nu(E)^2\rangle$ (\ref {eq:variance} ). The parameters
      are the same as in figure 1. The dashed
      line is a linear function of $E$ included
      for comparison.}
    \label{figure2}
  \end{center}
\end{figure}

We are unable to prove that the variance (\ref {eq:variance})
differs from zero, and that its $E$ dependence follows the
behavior expected from the numerical simulations.  However, we can
provide heuristic arguments which explain the systematic trends
observed in the data. The underlying assumptions are that

\noindent
\textit{I.} The number of nodal domains for individual
representing
vectors $\hat{\nu}[{\bf q}] = 2\sum_{r=1}^n |q_r|$
(\ref{eq:nutorus}) fluctuate independently about their mean
(\ref{eq:nuav}).

\noindent
\textit{II.} The fluctuations in $\delta
\nu_{Q^+}(E)$ and $\delta \nu_{Q^-}(E)$ are also independent.
(This is supported by the observation made in point {\textit{iii.}} of
the preceding section,  that only a fraction of the representation
vectors of the two Gram matrices can be transformed to each other
by a single rotation.)

\noindent
The first assumption together with the
central limit theorem  leads to the conclusion that
\begin{equation}
\left \langle  \left( \nu_{Q }(E)-\langle\nu_{Q
}(E)\rangle\right)^2\right \rangle \propto \frac{E}{\langle
 g_Q(E)\rangle}\ .
 \label{eq:estimate}
\end{equation}
The factor $E$ in the numerator comes because the nodal count
scales as $E^{\frac{1}{2}}$. The denominator $\langle
g_Q(E)\rangle$ is the mean number of integer points on the
ellipsoid. This behavior is consistent with recent number
theoretical estimates of the rate of convergence of the
fluctuations of the distribution of representations to the uniform
limit \cite {Rudnick}. Using the second assumption, we conclude
that
\begin{equation}
\left \langle  \left( \delta \nu_{Q }(E)\right )^2 \right \rangle
\propto \frac{E}{\langle
 g_Q(E)\rangle}\
 \label{eq:main}
\end{equation}
Thus, for \emph{rational} $Q^{\pm}$ where $\langle g_Q(E)\rangle
\propto E $, the variance is independent of $E$, whereas for
\emph{irrational} $Q^{\pm}$ the variance is expected to vary
linearly with $E$. This behavior is indeed observed in the
numerical simulations.

Another numerical test which supports the validity of the
above analysis, consisted in computing the fluctuations of the
integrated nodal counts
\begin{equation}
N_Q(E) = \frac{1}{\mathcal{N}(E)}\left ( \sum_{{\bf q} \in
\mathbb{Z}^n : {\bf q} \cdot Q {\bf q}\le E} \ \hat{\nu} [{\bf q}]
\right )\ ,  \label{eq:integrated}
\end{equation}
where $\mathcal{N}(E)\sim E^2$ is the spectral counting function
$\mathcal{N}(E)= \sum_{e\le E} g_Q(e)$. The difference between
integrated nodal counts for the isospectral pair can be written as
\begin{equation}
  \Delta N(E) =  N_{Q^{+}}(E)- N_{Q^{-}}(E)  =
  \frac{1}{\mathcal{N}(E)}\left(
    \sum_{{\bf q} \in V_+} \ \hat{\nu} [{\bf q}] -
    \sum_{{\bf q} \in V_-} \ \hat{\nu} [{\bf q}] \right),
 \label{eq:estimateN}
\end{equation}
where $V_\pm:=\{ {\bf q} \in \mathbb{Z}: {\bf q} \cdot Q_\pm {\bf q}\le E
\; \mathrm{and}\;
{\bf q} \cdot Q_{\mp} {\bf q}> E \}$. Here, we made use of the exact
cancellations of the contributions of ${\bf q}\in \mathbb{Z}$ in the
intersection of the two ellipsoids ${\bf q} \cdot Q_+ {\bf q}\le E$ and
${\bf q} \cdot Q_- {\bf q}\le E$. There are further exact
cancellations since $\hat{\nu}[{\bf q}]$ is invariant under reflections
$q_i \rightarrow - q_i$ and due to the symmetries
between the two ellipsoids.
Assuming that uncorrelated contributions to $\Delta N(E)$ stem
from a thin layer at the surface of the two ellipsoids while bulk
contributions cancel we have a sum over $s \sim |{\bf q}|^3= E^{3/2}$
independent contributions of order $\hat{\nu}[{\bf q}] \sim E^{1/2}$
that vanishes in the mean. For the variance this assumption leads to
$\langle N_Q(E)^2 \rangle \sim \frac{s \hat{\nu}[{q}]^2}{\mathcal{N}(E)^2}
\sim E^{-3/2}$ in accordance with our numerical analysis (see figure 3).
Note, that this result is valid for both, rational and irrational tori.
\begin{figure}
  \begin{center}
    \includegraphics[width=0.8\linewidth]{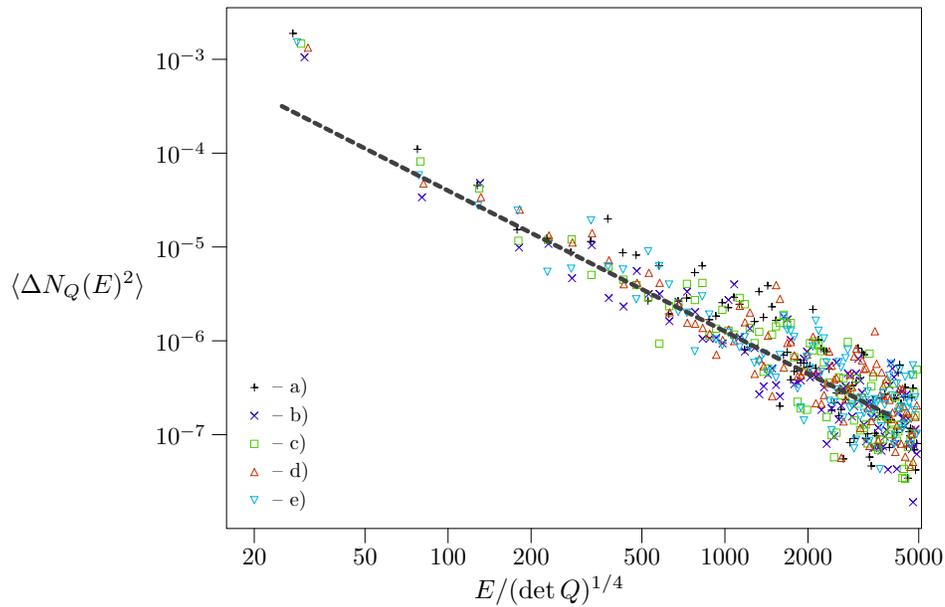}
    \caption{
      Double logarithmic plot of the squared nodal count
      difference
      $\langle \Delta N(E)^2 \rangle$ (smoothed by taking an average
      over a small spectral interval $\Delta E$).
      The parameters are the same as for figure 1.
      The dotted line shows the theoretical prediction $E^{-3/2}$.
    }
    \label{figure3}
  \end{center}
\end{figure}

\section {Summary and conclusions} \label{sec:sum}

The heuristic arguments as well as the numerical evidence
collected above, support  our conjecture that isospectral flat
tori can be distinguished by studying the fluctuations of their
nodal counts. A formal proof is still lacking. At the same time,
the work indicates a hitherto unnoticed  link between
isospectrality and nodal counts. We hope that the present work
will induce more research effort in this direction.

The nodal count  studied here is a particular function of the
representation vectors whose fluctuations distinguish between the
isospectral tori. Are there any other functions which are
sensitive to these differences? We investigated this question to
some extent, and were not able to find a simple criterion which
the function should satisfy in order to resolve the spectral
ambiguity. The nodal count was not chosen arbitrarily, and it is
rewarding that it does have the required property.

As was mentioned above, there exist several other known families
of flat tori beyond the ones discussed here at length. The way of
reasoning proposed above should apply to these cases as well.
Exploratory work in this direction shows that the nodal sequences
of isospectral pairs are different for the cases that were tried,
but a systematic study gets prohibitively time consuming as the
dimension increases.

The conjecture that nodal counts resolve isospectrality is now
being tested for a different class of operators - the
Schr\"odinger operators on quantum graphs \cite{shapiraus}.
Preliminary results support the validity of the conjecture also
for this class of operators which are quite different from the
flat tori discussed here.

\section {Acknowledgments}
This work was supported by the Minerva Center for non-linear
Physics, the Einstein (Minerva) center at the Weizmann Institute
and by  ISF and GIF research grants. NS acknowledges a post
doctoral fellowship from the European Network on {\it Mathematical
aspects of Quantum Chaos} which supported his stay at the Weizmann
Institute. US acknowledges support from the  EPSRC grant
GR/T06872/01. We would like to thank Professor Zeev Rudnick for
several crucial comments and discussions.

\appendix

\section {The mean nodal counts of isospectral tori}
\label{appendix1}
To demonstrate the equality of the smooth parts consider the simpler quantity
\begin{equation}
\tau_Q =  \int_{\mathbb{S}^3} {\rm d}\bomega \sum_{i} |\omega_{i}|
\cdot  \frac{1}{{(\bomega \cdot Q \cdot \bomega)}^{\eta}} \label
{eq:nuavPosOrth} \,
\end{equation}
with $\eta$ some power e.g. $5/2$. The goal is to show
$\tau_{Q^{+}} = \tau_{Q^{-}}$. From the diagonalization
(\ref{eq:tpm}):
\begin{equation}
Q^{\pm} = (T^{\pm})^{\sf T} D \, T^{\pm}
\end{equation}
make a change of variables
\begin{equation}
\mathbf{o} = T^{\pm} \cdot \bomega \,.
\end{equation}
This preserves the measure as $T^{\pm}$ are orthogonal and the
denominators are seen to agree as $D$ is the
same for the two lattices. The numerator becomes
\begin{equation}
\sum_{i} |\omega_{i}| = \sum_{i,j} |(T^{\pm})^{\sf T}_{i j} o_{j}|
= \sum_{i,j} | T^{\pm}_{j i} o_{j}| \,.
\end{equation}
By explicit calculation we get for the $``+"$ lattice:
\begin{eqnarray} \label{PlusSum}
\fl \sqrt{12} \, \sum_{i} |\omega_{i}|& = |o_{1} + o_{2} + o_{3} -
3 o_{4}| + |o_{1} - o_{2}- 3 o_{3} - o_{4}|   \\ \nonumber &+
|o_{1} + 3 o_{2} - o_{3} + o_{4}| + |3 o_{1} - o_{2} + o_{3} +
o_{4}|\ ,
\end{eqnarray}
and similarly for the $``-"$ lattice,
\begin{eqnarray} \label{MinSum}
\fl \sqrt{12} \, \sum_{i} |\omega_{i}|& = |o_1 - o_2 + 3 o_3 -
o_4|  + |o_1 - 3 o_2 - o_3 + o_4|  \\ \nonumber
  &+ |-3 o_1 - o_2 + o_3 + o_4| + |o_1 + o_2 + o_3 + 3 o_4 | \, .
\end{eqnarray}
Each absolute value for one lattice can be made to correspond  to
the other under the change of $o_{i}$ to $ -o_{i}$. For instance
the first term of the $``+"$  lattice matches with the fourth of
the $``-"$ lattice. In the integral for $\tau$ such sign changes
preserve both the integration domain, the measure and the
denominator. Therefore when splitting the integral into four,
corresponding to the terms of (\ref{PlusSum},\ref{MinSum}), each
of these terms can be brought to an agreement.

\end{document}